\documentclass[
    ,final            
  ]
  {aipproc}

\layoutstyle{8x11double}

\newcommand{\beq}{\begin{equation}}
\newcommand{\eeq}{\end{equation}}
\newcommand{\ba}{\begin{array}}
\newcommand{\ea}{\end{array}}
\newcommand{\beqa}{\begin{eqnarray}}
\newcommand{\eeqa}{\end{eqnarray}}

\newcommand{\cO}{{\cal O}}
\newcommand{\dis}{\displaystyle}
\newcommand{\no}{\nonumber}
\newcommand{\op}{Q}

\def\appol#1#2#3{  {\it Acta Phys. Polon.~B}, {\bf #1}, #3 (#2)}
\def\npb#1#2#3{    {\it Nucl. Phys.~B}, {\bf #1}, #3 (#2)}
\def\npbproc#1#2#3{  {\it Nucl. Phys. Proc. Suppl.}, {\bf #1}, #3 (#2)}
\def\plb#1#2#3{    {\it Phys. Lett.~B}, {\bf #1}, #3 (#2)}

\def\prd#1#2#3{    {\it Phys. Rev.~D}, {\bf #1}, #3 (#2)}

\def\prl#1#2#3{    {\it Phys. Rev. Lett.}, {\bf #1}, #3 (#2)}

\def\rmp#1#2#3{    {\it Rev. Mod. Phys.}, {\bf #1}, #3 (#2)}
\def\zpc#1#2#3{    {\it Z. Phys.~C}, {\bf #1}, #3 (#2)}
\def\ijmpa#1#2#3{  {\it Int. J. Mod. Phys.~A}, {\bf #1}, #3 (#2)}
\def\mpla#1#2#3{   {\it Mod. Phys. Lett. ~A}, {\bf #1}, #3 (#2)}

\def\epjc#1#2#3{   {\it Eur. Phys. J.~C}, {\bf #1}, #3 (#2)}
\def\jhep#1#2#3{   {\it JHEP}, {\bf #1}, #3 (#2)}

\def\ibid#1#2#3{{\bf #1}, #3 (#2)}

\begin{document}
\title{Theory of radiative and rare $B$ decays}
\author{Gino Isidori \\ }{
address={INFN, Laboratori Nazionali di Frascati, Via E. Fermi 40, I-00044 Frascati, Italy}}

\begin{abstract}
We present a concise theoretical overview of radiative and rare $B$ decays
mediated by flavour-changing neutral-current transitions 
of the type $b \to s(d) \gamma$ and $b \to s(d) \bar \ell \ell$. 
\end{abstract}

\maketitle

\section{Introduction}

Thanks to the efforts of $B$ factories, the exploration 
of the mechanism of quark-flavour mixing is now entering 
a new interesting era. The precise measurements 
of mixing-induced CP violation and tree-level allowed
semileptonic transition have provided an important
consistency check of the SM, and a precise determination 
of the Cabibbo-Kobayashi-Maskawa (CKM) matrix. 
The next goal is to understand if there is still 
room for new physics (NP) or, more precisely,
if there is still room for new sources of flavour symmetry 
breaking close to the electroweak scale.
From this perspective, radiative and rare $B$ decays
mediated by flavour-changing neutral current (FCNC) amplitudes 
represent a fundamental tool (see e.g. Ref.~\cite{rev}). 

Beside the experimental sensitivity, 
the conditions which allow to perform significant NP searches 
in rare decays can be summarized as follows:
i) decay amplitude dominated by electroweak dynamics,
and thus enhanced sensitivity to non-standard contributions;
ii) small theoretical error within the SM, or good control 
of both perturbative and non-perturbative corrections.
In the rest of this talk we shall analyze at 
which level these conditions are satisfied in various decay 
modes.

\section{Inclusive FCNC $B$ decays}

Inclusive rare $B$ decays such as $B\to X_s\gamma$,
$B\to X_s \ell^+ \ell^-$ and $B\to X_s\nu\bar\nu$
are the natural framework for high-precision
studies of FCNCs in the $\Delta B=1$ sector \cite{Hurth_rev}. 
Perturbative QCD and heavy-quark expansion 
form a solid theoretical framework to describe 
these processes: inclusive hadronic rates are related 
to those of free $b$ quarks, calculable in perturbation 
theory, by means of a systematic expansion in 
inverse powers of the $b$-quark mass.

The starting point of the perturbative partonic calculation 
is the determination of a low-energy effective
Hamiltonian, renormalized at a scale $\mu={\cal O}(m_b)$,
obtained by integrating out the heavy degrees 
of freedom of the theory. For $b\to s$ transitions 
--within the SM-- this can be written as
\beq
{\cal H}_{\rm eff}\! =\! - \frac{G_F}{\sqrt{2}}  V_{t s}^\ast  V_{tb}  
\sum_{i=1}^{10,~\nu}  C_i(\mu)  Q_i  + {\rm h.c.} \label{eq:he_DB}
\eeq 
where $\op_{1 \ldots 6}$ are four-quark operators, 
$Q_8$ is the chromomagnetic operator and 
\beqa
\op_7           &=& \dis\frac{e}{4 \pi^2} \bar{s}_L \sigma_{\mu \nu} 
                   m_b b_R F^{\mu \nu}  \no \\
\op_9           &=& \dis\frac{e^2}{4 \pi^2} \bar{s}_L \gamma^\mu b_L 
                   \bar{\ell} \gamma_\mu \ell  \no \\
\op_{10}        &=& \dis\frac{e^2}{4 \pi^2} \bar{s}_L \gamma^\mu b_L 
                   \bar{\ell} \gamma_\mu \gamma_5 \ell  \no \\
\op_{\nu}         &=& \dis\frac{e^2}{4 \pi^2 s_{w}^2} 
\bar{s}_L\gamma^\mu b_L~\bar{\nu}_L \gamma_\mu \nu_L \label{eq:4ops}
\eeqa
are the leading FCNC electroweak operators. 
Within the SM, the coefficients of all the operators 
in Eq.~(\ref{eq:4ops}) receive a large non-decoupling 
contribution from top-quark loops
at the electroweak scale. But the $m_t$ 
dependence is not the same for the four operators, 
reflecting a different $SU(2)_L$-breaking structure,
which can be affected in a rather different 
way by new-physics contributions \cite{Bertolini:1986tg,MFV}.

The calculation of the rare decay rates then involves three distinct 
steps: i) the determination of the initial conditions of the Wilson 
coefficients at the electroweak scale; ii) the evolution by means of 
renormalization-group equations (RGEs) of the $C_i$ down 
to $\mu={\cal O}(m_b)$; iii) the evaluation of the hadronic matrix 
elements of the effective operators at $\mu={\cal O}(m_b)$,
including both perturbative and non-perturbative QCD corrections.  
Each of the three steps must be taken to matching orders of accuracy 
in powers of the strong coupling constants $\alpha_s$ and 
of the large logs generated by the RGE running.
The interesting short-distance (electroweak) dynamics that 
we would like to test enters only in the first step;  
the following two steps are fundamental ingredients 
to reduce and control the theoretical error. 

\begin{figure}[t]
  \includegraphics[height=.14\textheight]{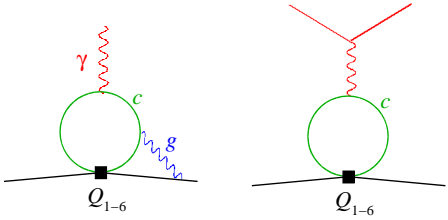}
\caption{Representative diagrams for the mixing of 
four-quark operators into $Q_7$ (left) and $Q_9$ (right).}
\label{fig:Q16}
\end{figure}

The first two steps (initial conditions and RGEs) are 
process independent and are common also to exclusive 
modes. Nonetheless, the organization of the leading-log (LL)
series is not the same for the three underlying 
partonic processes, or the four operators in Eq.~(\ref{eq:4ops}):

$\underline{b \to s \gamma}$. Here only $Q_7$ 
has a non-vanishing matrix element at the tree level.
The large logarithms generated by mixing of four-quark operators 
into $Q_7$ (see Fig.~\ref{fig:Q16}) play a very important role and 
enhance the partonic rate by a factor of almost three \cite{Bertolini}.
Since this mixing vanishes at the one-loop level, 
a full treatment of QCD corrections beyond lowest order 
is a rather non-trivial task. This has been achieved 
already a few years ago, thanks to the joint effort 
of many authors (see e.g. Ref.~\cite{Hurth_rev,BM_rev} and references therein),
and is nowadays a rather mature subject.
All the ingredients of the partonic calculation 
have been cross-checked by more than one group. 
In particular, very recently an independent confirmation 
of the three-loop mixing of 
$Q_7$ and $Q_{1 \ldots 6}$ \cite{Misiak} 
-- till few months ago the only piece of the calculation performed 
by one group only -- has been presented \cite{Paolo_new}. 

$\underline{b \to s \ell^+\ell^-}$. The 
three operators with non-vanishing tree-level 
matrix elements are $Q_7$, $Q_{9}$ and $Q_{10}$.
Similarly to  $Q_7$, QCD corrections are very 
important also for $Q_{9}$. Since $Q_9$ 
mixes with four-quark operators already 
at the one-loop level (see Fig.~\ref{fig:Q16}), 
the organization of the LL series for  
$b \to s \ell^+\ell^-$ is different than in 
$b \to s \gamma$: the NLL level 
is much simpler (no three-loop mixing involved),
but less precise. An accuracy below 
the $10\%$ level on the decay rate (a precision similar 
to the NLL level in $b \to s \gamma$) is reached here only 
with a NNLL calculation. All the missing ingredients to reach this goal 
has finally become available. In particular, NNLL
initial conditions and anomalous dimension matrix 
can be found in \cite{BobethM} and \cite{Paolo_new},
respectively. 

It is worth to stress that the impact of QCD corrections 
is very limited in the axial-current operator $Q_{10}$.
This operator does not mix with four-quark operators 
and is completely dominated by short-distance 
contributions. Together with $Q_\nu$, $Q_{10}$
belongs to the theoretically clean $\cO(G_F^2)$ hard-GIM-protected  
part of the effective Hamiltonian.
Thus observables more sensitive to $Q_{10}$, 
such as  the forward-backward (FB) lepton asymmetry
in $b \to s \ell^+\ell^-$, have a 
reduced QCD uncertainty and a stronger sensitivity 
to possible non-standard phenomena.

$\underline{b \to s \nu \bar\nu}$. In this case 
only $Q_\nu$ is involved. Similarly to $Q_{10}$, 
QCD corrections play a very minor role 
since there is no mixing with  four-quark operators. 
As a result, the only non-trivial step of the 
perturbative calculation for $b \to s \nu \bar\nu$
decays is the determination of the initial condition 
of $C_\nu$ at the electroweak scale: 
this is known with a precision around $1\%$ 
within the SM \cite{bsnn}.

\smallskip
These two processes-independent steps of the calculation,
namely the determination of the effective Hamiltonian
renormalized at a low scale $\mu={\cal O}(m_b)$,
can easily be transferred from the $b\to s$ case to the $b \to d$ one.
The only difference is the richer CKM structure of the 
$b \to d$ Hamiltonian, with two independent 
non-negligible terms ($ V_{t d}^\ast  V_{tb}$ and $V_{u d}^\ast  V_{ub}$).

The situation is very different for the last step of the calculation,
namely the evaluation of the hadronic matrix elements. The latter
strongly depend on the specific process and the specific observable 
we are interested in (e.g.~fully inclusive rate 
or differential distribution). 
%
In the following we shall review the results of 
this step (and thus the final numerical predictions) 
for some of the most interesting 
$b\to s$ observables.

\subsection{$B \to X_s \gamma\ $ [{\em the most effective ``NP killer''}]}
The inclusive $B \to X_s \gamma$ rate is 
the most precise and clean short-distance information that 
we have, at present, on $\Delta B=1$ FCNCs. 
Combining the precise measurements by ALEPH,
BaBar, Belle and CLEO, the world average reads \cite{Nakao}
\beq
{\cal B}( B \to X_s \gamma )^{\rm exp} = (3.34 \pm 0.38) \times 10^{-4}
\label{eq:bsg_exp}
\eeq

On the theory side, the NLL partonic calculation of the 
matrix elements, performed first in Ref.~\cite{Greub} for the 
leading terms, has recently been cross-checked and completed in Ref.~\cite{BM_new}.
Perturbative corrections due to higher-order electroweak effects 
have also been analyzed (see Ref.~\cite{Paolo_bsg_em} and
references therein).

Non-perturbative $1/m_b$ corrections 
are well under control in the total rate. In particular,  
$\cO(1/m_b)$  corrections vanish in the ratio 
$\Gamma( B \to X_s \gamma )/\Gamma(B \to X_c \ell \nu )$,
and the $\cO(1/m^2_b)$ ones are known and amount to few per cent \cite{Falk}.
Also non-perturbative effects associated to charm-quark loops 
have been estimated and found to be very small \cite{Voloshin,LLW,BIR}.
The most serious problem of non-perturbative origin is related 
to the (unavoidable) experimental cut in the photon energy 
spectrum that prevents the measurement from being fully 
inclusive \cite{LLW,KaganN}. With the present cut by CLEO $E_\gamma > 2.0$~GeV
\cite{CLEO_bsg}, this uncertainty is smaller but non-negligible 
with respect to the error of the perturbative calculation.
The latter is around $10\%$ and its main source is the 
uncertainty in the ratio $m_c/m_b$ that enters through 
charm-quark loops \cite{MisiakG}. 

According to the detailed  
analysis of theoretical errors presented in Ref.~\cite{MisiakG},
the SM expectation~is 
\beq
{\cal B}(B \to X_s \gamma)^{\rm SM} =  (3.73 \pm 0.30) \times 10^{-4}~,
\label{eq:bsg_MG}
\eeq
in good agreement with Eq.~(\ref{eq:bsg_exp}). 
It must be stressed that the overall scale dependence is very small:
for $\mu \in [m_b/2,2 m_b]$ the central value moves 
by about $1\%$. The error in Eq.~(\ref{eq:bsg_MG}) is 
an educated guess about the size of possible NNLL terms.
In particular, the largest source of uncertainty is 
obtained by the variation of 
${\overline m}_c(\mu)/m_b^{\rm pole}$  for $\mu \in [m_c , m_b]$.  
A critical discussion about the error in Eq.~(\ref{eq:bsg_MG}),
with alternative more conservative estimates,
can be found in Ref.~\cite{Porod}.

The comparison between theory and experiments in 
${\cal B}(B \to X_s \gamma)$ is a great success of the 
SM and has led us to derive many significant bounds on possible 
new-physics scenarios. For instance, the ${\cal B}(B \to X_s \gamma)$  
constraint is one of the main obstacles to build consistent 
models that predict a sizable difference between 
${\cal A}_{\rm CP}(B \to \phi K_S)$ and 
${\cal A}_{\rm CP}(B \to \psi K_S)$. This constraint can 
be avoided (see  e.g. Ref.~\cite{ciuchini} and references there in), 
but the resulting models require a considerable amount of 
fine tuning. By far more natural are the so-called MFV models \cite{MFV}, 
where ${\cal A}_{\rm CP}(B \to \phi K_S) \approx {\cal A}_{\rm CP}(B \to \psi K_S)$
and deviations from the SM in ${\cal B}(B \to X_s \gamma)$ do not 
exceed the $10\%$--$30\%$ level \cite{MFV}. 
Improved measurements of ${\cal B}(B \to X_s \gamma)$ 
are certainly useful to further constrain this possibility.
However, since the experimental 
error has reached the level of the theoretical one,
it will be very difficult to clearly identify possible 
deviations from the SM, if any, in this observable.

Hopes to detect new-physics signals are still 
open through the CP-violating asymmetry 
\beq
\Delta \Gamma_{\rm CP} (B \to X_s \gamma ) =
\frac{\Gamma (B \to X_s \gamma ) - \Gamma (\bar{B} \to X_s \gamma ) }
{\Gamma (B \to X_s \gamma ) + \Gamma (\bar{B} \to X_s \gamma ) }~.\
\label{eq:a_s_cp}
\eeq
This is expected to be below $1\%$ within the SM \cite{Porod,KaganN2},
but could easily reach $\cO(10\%)$ values beyond the SM, even 
in the absence of large effects in the total $B \to X_s \gamma$
rate. This is indeed one of the main expectations in models 
with sizable NP effects in ${\cal A}_{\rm CP}(B \to \phi K_S)$.
The present measurement of $\Delta \Gamma_{\rm CP} (B \to X_s \gamma )$
is consistent with zero \cite{Nakao}, but the  sensitivity
is still one order of magnitude above the SM level.

\subsection{$B\to X_s\ell^+\ell^-\ $ [{\em the present frontier}] }
Both Belle \cite{Belle_bsll} and 
BaBar \cite{Babar_bsll} 
have recently announced a clear evidence ($\approx 5 \sigma$) 
of the $B\to X_s\ell^+\ell^-$ decay. The two results are 
compatible and are both based on a semi-inclusive analysis
(the hadronic system is reconstructed from a kaon 
plus $0$ to $4$ pions, with at most one $\pi^0$).
Their combination \cite{Nakao}
\beq
{\cal B}( B \to X_s \ell^+\ell^- )^{\rm exp} = (6.2 \pm 1.1 {}^{+1.6}_{-1.3} ) \times 10^{-6}~.
\label{eq:bsll_exp}
\eeq
represents a very useful new piece of information 
about $\Delta B=1$ FCNCs, with considerable margin 
of improvement in the near future. 

In principle, these decays offer a phenomenology 
reacher than $B\to X_s\gamma$, with more than one 
interesting observable. The joint effort of several
groups has recently allowed to evaluate at the NNLL level
all the matrix element necessary for the two main 
kinematical distributions: the dilepton 
spectrum \cite{Asa1,Adrian2,Paolo_bsll}
and the lepton FB asymmetry \cite{Adrian1,Asa2}. 

\begin{figure}[t]
\includegraphics[height=.26\textheight]{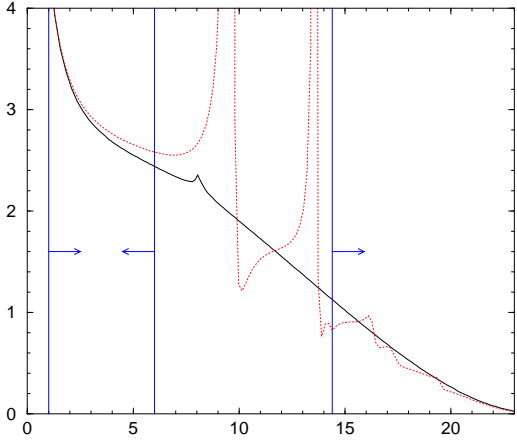}
\caption{Dilepton spectrum of the inclusive  
$B \to X_s e^+ e^-$ decay within the SM. Vertical axis:
$d{\cal B}(B\to X_s e^+ e^-)/dq^2$ in units of 
$10^{-7}\times{\rm GeV}^{-2}$; horizontal axis:  $q^2$ in ${\rm GeV}^2$.
The dotted line denotes the NNLL pure perturbative result, the full line 
includes an estimates of the non-perturbative $c\bar{c}$ 
effects \cite{Adrian2}.}
\label{fig:bsee}
\end{figure}

In addition to the non-perturbative corrections due to the finite $b$ quark mass, 
$B\to X_s \ell^+\ell^-$ transitions suffer from specific 
non-perturbative effects due to long-lived $c\bar{c}$ intermediate states
($B \to X_s c\bar{c} \to  X_s \ell^+\ell^-$).
The heavy-quark expansion, which allow to evaluate
the $\Lambda_{\rm QCD}/m_b$ terms, is rapidly convergent and leads 
to small corrections for {\em sufficiently inclusive} observables
\cite{AHHM,BI}. A consistent treatment 
of the second type of effects requires 
{\em kinematical cuts} in order to avoid the large 
non-perturbative background of the narrow $c\bar{c}$ resonances
(see Fig.~\ref{fig:bsee}).
These two requirements are somehow in conflict \cite{Adrian2,BI}; 
nonetheless, we can identify two {\rm perturbative windows},
defined by:
\beqa
&& q^2 \equiv  M^2_{\ell^+ \ell^-} ~ \in ~ [1~\mbox{GeV}^2,~6~\mbox{GeV}^2] \quad \mbox{(low)}~, \no \\
&& q^2 > 14.4 \ \mbox{GeV}^2  \qquad\qquad\qquad\qquad   \mbox{(high)}~, \no
\eeqa
where reliable predictions can be performed \cite{Adrian2}. 
It is worth to emphasize that the two regions  
have complementary virtues and disadvantages:
\begin{itemize}
\item
{\em Virtues of the low-$q^2$ region:} reliable $q^2$ spectrum; 
small $1/m_b$ corrections; sensitivity to the interference of $C_7$
and $C_9$; high rate.
\item
{\em Disadvantages of the low-$q^2$ region:} difficult to perform 
a fully inclusive measurement (severe cuts on the dilepton 
energy and/or the hadronic invariant mass); long-distance effects 
due to processes of the type $B \to \Psi X_s \to  X_s + X^\prime \ell^+\ell^-$ 
not fully under control; non-negligible scale 
and $m_c$ dependence.
\item
{\em Virtues of the high-$q^2$ region:} negligible scale 
and $m_c$ dependence due to the strong sensitivity to $|C_{10}|^2$;
easier to perform a fully inclusive measurement (small 
hadronic invariant mass); negligible long-distance effects of the type 
$B \to \Psi X_s \to  X_s + X^\prime \ell^+\ell^-$.
\item
{\em Disadvantages of the high-$q^2$ region:} $q^2$ spectrum  
not reliable (only the integrated rate can be predicted); 
sizeable $1/m_b$ corrections (effective expansion in 
$1/(m_b-\sqrt{q_{\rm min}})$ \cite{Adrian2}); low rate.
\end{itemize}
Given this situation, we believe that future experiments should 
try to measure the branching ratios in both regions and 
report separately the two results. These two measurements 
are indeed affected by different systematic uncertainties 
(of theoretical nature) and provide a different 
short-distance information. The NNLL SM predictions
for the two clean windows \cite{Adrian2},
\beqa
{\cal B}(B\to X_s \ell^+\ell^-)_{\rm low}^{\rm SM} &=& (1.63 \pm 0.20) \times 10^{-6}~, \no \\
{\cal B}(B\to X_s \ell^+\ell^-)_{\rm high}^{\rm SM} &=& (4.04 \pm 0.78) \times 10^{-7}~, \quad 
\label{eq:SM_bsll}
\eeqa
are still affected by a considerable error; however, in both
cases the uncertainty is mainly of parametric nature and could 
be substantially improved in the future. In particular, the 
large error in the high-$q^2$ region is mainly due to the  uncertainty 
in the relation between the physical $q^2$ interval and the corresponding 
interval for the partonic calculation (i.e. the uncertainty in the 
relation between $m_b$ and the physical hadron mass), which can be improved 
with better data on charged-current semileptonic modes. 
According to the recent analysis of Ref.~\cite{Paolo_bsll}, 
both results in (\ref{eq:SM_bsll}) should be decreased by 
$\approx 4\%$ to take into account the leading electroweak 
corrections. 

The two results in (\ref{eq:SM_bsll}) cannot be 
directly confronted with (\ref{eq:bsll_exp}), which includes 
an extrapolation to the full $q^2$ spectrum. The 
updated SM expectation for this extrapolated branching 
ratio is  $(4.6 \pm 0.8) \times 10^{-6}$  \cite{Adrian2}
(in agreement with the previous estimate of Ref.~\cite{Ali_extrap}),
and it is consistent with the present experimental world average.
We stress that this prediction is already saturated by irreducible 
theoretical errors and, contrary to the results in (\ref{eq:SM_bsll}), 
is very difficult to improve it further. 

\smallskip 

As anticipated, some of the most interesting short-distance 
tests in ${\cal B}(B\to X_s \ell^+ \ell^-)$ decays 
can be performed by means of the FB asymmetry of the 
dilepton distribution:
\beqa
{\cal A}_{\rm FB}(q^2) &=& \frac{1}{d {\cal B} ( B\to X_s \ell^+\ell^-) /d q^2  }
  \int_{-1}^1 d\cos\theta_\ell ~ \no \\
&& 
\frac{d^2 {\cal B} ( B\to X_s \ell^+\ell^-)}{d q^2  ~ d\cos\theta_\ell}
\mbox{sgn}(\cos\theta_\ell)~, \quad
\label{eq:asdef}
\eeqa
where  $\theta_\ell$ is the angle between 
$\ell^+$ and $B$  momenta in the dilepton 
centre-of-mass frame. Here the SM 
predict a zero for $s_0 = q_0^2/m^2_b = 0.162 \pm 0.008$ 
\cite{Adrian1,Asa2}: a very precise prediction which could 
easily be modified beyond the SM, even in absence of 
significant non-standard effects on the total rate.

\section{Exclusive modes}

On general grounds, theoretical predictions for 
exclusive FCNC decays are more difficult. 
The simplest cases are processes 
with at most one hadron in the final state. 
Here there has been a substantial progress in the last few years, 
both by means of analytic approaches \cite{Stewart} and by means 
of Lattice QCD \cite{Aida}, but still the overall theoretical 
uncertainty is around $20\%$ at the amplitude level.
The largest source of uncertainty is typically 
the normalization of the hadronic form factors, 
an error that can be substantially reduced in appropriate 
ratios or differential distributions. These type 
of observables become particularly interesting in 
channels where,  because of irreducible experimental 
problems, the short-distance 
amplitude  cannot be extracted from corresponding 
inclusive modes. Two of such examples are the ratio 
\beq
R_{\gamma }(\rho/K^*) = \frac{ {\cal B} (B\to \rho \gamma) }{ {\cal B} (B\to K^* \gamma) }~,
\eeq
and the normalized FB asymmetry in $B\to K^* \ell^+\ell^-$. 

\smallskip

\begin{figure}[t]
\includegraphics[height=.16\textheight]{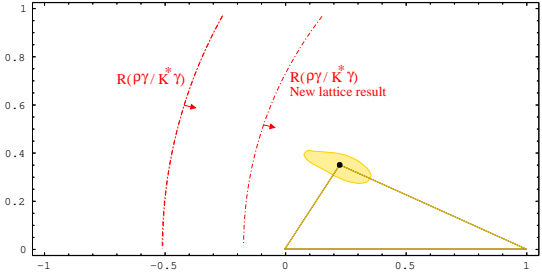}
\caption{Bounds on the $\rho$--$\eta$ plane from the 
BaBar bound on $R_{\gamma }(\rho/K^*)$: the two curves correspond 
to different values of $\zeta$ \cite{Hurth_Lunghi}. } 
\label{fig:Brhogamma}
\end{figure}

The ratio $R_{\gamma }(\rho/K^*)$ is one of the most promising tool 
to extract short-distance properties about the $b\to s \gamma$ amplitude. 
On the experimental side, the combination of the bounds on charged 
and neutral channels, in the isospin limit, leads to 
$R_{\gamma }(\rho/K^*) <0.047$ at 90\% C.L. \cite{Babar_rhog}. 
On the theory side, the $B \to V \gamma$ amplitudes which determine 
this ratio have been analyzed beyond naive factorization by several authors 
\cite{Brhogamma,Hurth_Lunghi,Martin}.
Within the SM one can write 
\beq
R_{\gamma }(\rho/K^*) = \left| \frac{V_{td}}{V_{ts}} \right|^2 \frac{ (m^2_B -m^2_{rho} )^3}{ 
(m^2_B -m^2_{K^*} )^3 }\zeta^2 (1-\Delta R)~,
\eeq
where $\zeta$ denotes the ratio of 
the form factors at $q^2=0$ in the $m_b\to \infty$ limit,
and $\Delta R$ the additional $SU(3)$ (and isospin) 
breaking due to $1/m_b$ and  $\cO(\alpha_s)$ corrections. 
The largest source of uncertainty is $\zeta$: 
according to the light-cone sum rule estimate $\zeta=0.76\pm 0.10$
\cite{Hurth_Lunghi}, the present experimental bound 
on $R_{\gamma }(\rho/K^*)$ is about twice the SM expectation. 
However, preliminary Lattice results indicate a 
larger value, $\zeta=0.91\pm 0.08$ \cite{Becirevic}, 
which would imply 
more stringent bounds on  $|V_{td}/V_{ts}|$. The two 
curves in Fig.~\ref{fig:Brhogamma} summarizes 
the present status.

\smallskip

\begin{figure}[t]
  \includegraphics[height=.2\textheight]{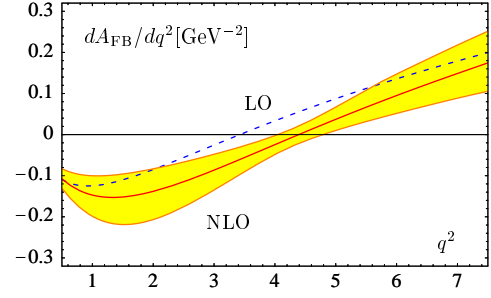}
\caption{Zero of the forward-backward asymmetry in
$B^- \to K^{*-} \ell^+\ell^-$ at LO and NLO. 
The band reflects all theo\-retical uncertainties from 
parameters and scale dependence combined \cite{Martin}. } 
\label{fig:martin}
\end{figure}

While the inclusive $B\to X_s \ell^+\ell^-$ rate 
is already accessible at the $B$ factories, 
a differential study of the inclusive FB asymmetry 
is beyond their present and near-future reach. More accessible 
from the experimental point of view is the FB asymmetry in $B\to K^* \ell^+\ell^-$ 
(defined as in Eq.~(\ref{eq:asdef}) with $X_s \to K^*$).
Assuming that the leptonic current has only a 
vector ($V$) or axial-vector ($A$) structure
(as in the SM), the FB asymmetry provides a direct measure of 
the $A$--$V$ interference. Indeed, at the lowest-order one can write
$$
{\cal A}_{\rm FB}(q^2)
 \propto 
  {\rm Re}\left\{  C_{10}^* \left[ \frac{q^2}{m_b^2} C_9^{\rm eff} 
    + r(q^2) \frac{m_b C_7}{m_B}  \right] \right\}~,
$$
where $r(q^2)$ is an appropriate ratio of $B\to K^*$
vector and tensor form factors \cite{burdman0}. 
There are three main features of this observable
that provide a clear and independent 
short-distance information: 
1)~The position of the zero of ${\cal A}_{\rm FB}(q^2)$ in the 
low-$q^2$ region (see Fig.~\ref{fig:martin}) \cite{burdman0}.
As shown by means of a full NLO calculation \cite{Martin}, 
the experimental measurement of $q^2_0$ could 
allow a determination of $C_7/C_9$ at the  $10\%$ level.
2)~The sign of ${\cal A}_{\rm FB}(q^2)$ around the zero.
This is fixed unambiguously in terms of the relative sign
of $C_{10}$ and $C_9$ \cite{BHI}: within the SM one 
expects ${\cal A}_{\rm FB}(q^2 > q^2_0) > 0$  for 
$|\bar B \rangle \equiv |b \bar d \rangle$ mesons.
3)~The relation ${\cal A}[\bar B]_{\rm FB}(q^2) = - {\cal A}[B]_{\rm FB}(q^2)$.
This follows from the CP-odd structure of ${\cal A}_{\rm FB}$
and holds at the $10^{-3}$ level within the SM \cite{BHI}, 
where $C_{10}$ has a negligible CP-violating phase.

\subsection{$B_{s,d} \to \ell^+\ell^-\ $ [{\em the future frontier}] }
The purely leptonic decays constitute a 
special case among exclusive transitions. Within the SM 
only the axial-current operator, $Q_{10}$, 
induces a non-vanishing contribution 
to these decays. As a result, the short-distance contribution
is not {\em diluted} by the mixing with four-quark operators.
Moreover, the hadronic matrix element involved is the simplest
we can consider, namely the $B$-meson decay constant
\beq
\langle 0 | \bar q \gamma_\mu \gamma_5 b | \bar B_q (p) \rangle
= i p_\mu f_{B_q} 
\eeq
Reliable estimates of $f_{B_d}$  and $f_{B_s}$ are 
obtained at present from lattice calculations
and in the future it will be possible to cross-check
these results by means of the $ B^+ \to \ell^+ \nu$ rate.
Modulo the determination of $f_{B_q}$, the theoretical
cleanliness of $B_{s,d} \to \ell^+\ell^-$ decays
is comparable to that of the {\em golden modes}
$K_L \to \pi^0 \nu \bar{\nu}$ and
$B\to X_{s,d} \nu\bar\nu$.

The price to pay for this theoretically-clean amplitude is a strong helicity 
suppression for $\ell=\mu$ (and $\ell=e$), or the channels 
with the best experimental signature. Employing the 
full NLO expression  of $C_{10}$ \cite{bsnn},
we can write 
\beqa
&&{\mathcal B}( B_s \to \mu^+ \mu^-)^{\rm SM} = 3.1 \times 10^{-9} 
\left( \frac{|V_{ts}|}{0.04}  \right)^2 \no\\
&& \times \left( \frac{f_{B_s}}{0.21~\mbox{GeV}} \right)^2 \!\!
\left( \frac{\tau_{B_s}}{1.6~\mbox{ps}} \right)
\left( \frac{ m_t(m_t) }{166~\mbox{GeV} } \right)^{3.12} \quad
\label{eq:BrmmSM} \no \\ \no \\
&&\qquad\quad  \frac{ {\mathcal B}( B_s \to \tau^+ \tau^-)^{\rm SM} }{  
{\mathcal B}( B_s \to \mu^+ \mu^-)^{\rm SM} } = 215~.  \no
\eeqa
The corresponding $B_d$ modes are both 
suppressed by an additional factor $|V_{td}/V_{ts}|^2$ 
$=(4.0 \pm 0.8)\times10^{-2}$. The present experimental 
bound closest to SM expectations is the one obtained by 
CDF on $B_{s}\to \mu^+ \mu^-$ \cite{Nakao,CDF}:
$$
{\mathcal B}(B_{s}\to \mu^+ \mu^-) < 9.5 \times 10^{-7} \quad 
(95 \% \;\;  {\rm CL})~,
$$
which is still very far from the SM level. The latter will 
certainly not be reached before the LHC era.

As emphasized in the recent lit\-era\-ture \cite{Babu,Bobeth,nonMFV},
the purely leptonic decays of $B_s$ and $B_d$
mesons are excellent probes of several new-physics
models and, particularly, of scalar FCNCs. 
Scalar FCNC operators, such as $\bar b_R s_L \bar \mu_R \mu_L$, 
are present within the SM but are absolutely 
negligible because of the smallness 
of down-type Yukawa couplings. On the other hand,
these amplitudes could be non-negligible
in models with an extended Higgs sector.
In particular, within the MSSM, where two Higgs doublets are 
coupled separately to up- and down-type quarks, a strong 
enhancement of scalar FCNCs can occur 
at large $\tan\beta = v_u/v_d$ \cite{Babu}. 
This effect is very small in non-helicity-suppressed 
$B$ decays and in $K$ decays (because of the small 
Yukawa couplings), but could enhance $B\to \ell^+\ell^-$
rates by orders of magnitude. As pointed out in 
Ref.~\cite{Hiller_new}, $\cO(100)$ enhancements 
in ${\cal B}(B\to \ell^+\ell^-)$ correspond to $\cO(10\%)$
breaking of universality in  ${\cal B}(B\to K \mu^+ \mu^-)$ 
vs.~${\cal B}(B\to K e^+ e^-)$.
Therefore, the present search for $B\to \ell^+\ell^-$ at 
CDF is already quite interesting, even if the sensitivity 
is well above the SM level. In a long-term perspective, 
the discovery of such processes is definitely one of 
the most interesting items in the $B$-physics program 
of hadron colliders.

\section{Conclusions}
Rare FCNC decays of $B$ mesons provide a unique 
opportunity to perform high-precision studies 
of quark-flavour mixing. The $B \to X_s \gamma$ rate,
where both experimental and theoretical errors have reached 
a comparable level around $10\%$, represents 
the highest peak in our present knowledge of FCNCs. 
The lack of deviations from SM expectations in $\Gamma(B \to X_s \gamma)$
should not discourage the measurement of other clean and 
independent FCNC observables, such as the forward--backward 
asymmetry in $B \to X_s \ell^+\ell^-$ or the $B \to \ell^+ \ell^-$ rates. 
Even if new physics will first be discovered elsewhere,
the experimental study of these theoretically-clean observables 
would still be very useful to investigate the flavour structure 
of any new-physics scenario.

\subsection*{Acknowledgements}
I am grateful to the organizers of Beauty 2003 for the invitation 
and the financial support that allowed me to attend
this interesting conference. This work is partially 
supported by the EC-Contract HPRN-CT-2002-00311 (EURIDICE).

\bibliographystyle{aipproc}   

\end{document}